\documentclass[aip,jcp,onecolumn,12pt,superscriptaddress]{revtex4-2}

\usepackage{amsmath,amssymb,graphicx}
\usepackage{bm}
\usepackage[hidelinks]{hyperref}
\hypersetup{pdftitle={Activity-induced response amplification and instability near jamming in a Wishart-Ginibre model},
             pdfauthor={Hisao Hayakawa}}
\usepackage{comment}
\usepackage{color}

\begin{document}

\title{Activity-induced response amplification and instability near jamming in a Wishart--Ginibre model}

\author{Hisao Hayakawa}
\affiliation{Yukawa Institute for Theoretical Physics, Kyoto University, Kyoto 606-8502, Japan}
\email{hisao@yukawa.kyoto-u.ac.jp}

\date{\today}

\begin{abstract}
We study the linear mechanical response of a disordered contact network using an additive Wishart--Ginibre model: a passive Wishart operator is perturbed by an independent real Ginibre coupling of strength \(\alpha_0\). 
The observable is the mean-square static displacement response per degree of freedom, normalized by the variance of the applied isotropic forces, rather than a shear compliance. 
In the large-matrix limit, the Hermitized Dyson equation gives \(J=(1-q)^{-3}/[1-(\alpha_0/\alpha_c)^2]\) on the stable side,
where \(q<1\) is the Wishart aspect ratio and \(\alpha_c=(1-q)^{3/2}\). 
Thus the random coupling amplifies the response up to the onset of linear instability. 
At the unshifted isostatic point \(q=1\), the limiting spectrum enters the right half-plane for every nonzero \(\alpha_0\), so there is no stable isostatic susceptibility at nonzero activity. 
Direct matrix calculations show response amplification away from the threshold and a broadened stability crossover at finite size. 
Near marginality, a single singular value often carries much of the inverse trace, including in unstable realizations for which that trace is not a steady response.
\end{abstract}

\maketitle

\section{Introduction}

A disordered solid close to jamming can have a large displacement response even when its contacts are individually stiff.
In passive packings, marginal rigidity is associated with an excess of soft modes of the contact Hessian \cite{LiuNagel1998,OHern2003,Wyart2005,Franz2015}. 
In dense active materials, including self-propelled particles and confluent tissues, both persistent driving and structural rearrangements can change the mechanical response \cite{Henkes2011,Park2015,Bi2016,Atia2018}.
This raises a question that can be posed already at the level of a fixed contact network: does a non-reciprocal displacement coupling suppress the passive soft-mode response, or does it amplify the response and destabilize the reference state?

That question requires distinguishing propulsion from feedback. 
A position-independent self-propulsion force at fixed orientation does not contribute to the displacement Jacobian. Even when the
orientation fluctuates, a torque-free orientation process produces a forcing term rather than a displacement feedback
(Sec.~\ref{sec:active_janus_grains}). 
Explicitly non-reciprocal interactions or deformation-dependent internal variables can instead produce a nonsymmetric linear operator. 
Such a coupling motivates, but does not uniquely specify, the random-matrix model studied here.

We represent the passive elasticity by a Wishart matrix \(\mathcal H_0\), with aspect ratio \(q\) and a Marchenko--Pastur spectrum \cite{Marchenko1967}, and model the non-reciprocal feedback by an independent real Ginibre matrix \(\mathcal A\). 
This gives \(\mathcal M=-\mathcal H_0+\alpha_0\mathcal A\). 
The strength \(\alpha_0\) characterizes the effective matrix, not a calibrated microscopic propulsion force or tumbling rate. We ask how \(\alpha_0\) changes the stability of \(\mathcal M\) and the response to isotropic, time-independent body forces. These are distinct questions: the rightmost eigenvalues determine asymptotic linear stability, whereas the squared static response is governed by the
singular values. 
An algebraic inverse may therefore exist even when a constant force does not produce a steady response.

Our analysis specializes established results for additive deformations of Ginibre matrices and their squared resolvents
\cite{Girko1985,AFM2015,HoZhong2023} to a contact-network surrogate with a tunable passive edge. 
At fixed \(q<1\), the passive inverse-square moment is \((1-q)^{-3}\). 
The effective activity amplifies this response up to the bulk threshold \(\alpha_c=(1-q)^{3/2}\). 
For the frictionless contact counting \(q\simeq 2D/Z\), this gives, close to isostaticity,
\begin{equation}\label{def:alpha_c}
 \alpha_c:=\left(\frac{Z-2D}{Z}\right)^{3/2}
 \simeq \left(\frac{Z-2D}{2D}\right)^{3/2}.
\end{equation}
Thus, a specific prediction of the Wishart--Ginibre surrogate is that the stable interval of non-reciprocal coupling collapses with the
\(3/2\) power of the excess coordination. 
This prediction concerns the effective displacement Jacobian in passive spectral units, rather than a microscopic propulsion force; Sec.~\ref{sec:discussion_tests}
discusses how the corresponding effective coupling could be measured independently from a microscopic Jacobian.
Direct calculations test the stable-side prediction and separately track the fraction of stable samples and the share of the smallest singular value.

Sections~\ref{sec:active_janus_grains} and \ref{sec:RMT_tumor} introduce the physical setting and the ensemble calculation, respectively. 
Section~\ref{sec:direct_random_matrix_results} presents the numerical tests. 
The appendices give the Hermitized derivation, the forcing interpretation of \(J\), and the scope of the mechanical modeling assumptions.

%%%%%%%%%%%%%%%%%%%%%%%%%
\section{Microscopic model and effective active couplings}
\label{sec:active_janus_grains}
%%%%%%%%%%%%%%%%%%%%%%%%%

We consider a disordered packing of \(N\) frictionless spherical grains in \(D\) dimensions at packing fraction \(\phi\). 
We distinguish the microscopic dynamics of particles with self-propulsion from the statistical model used below for a non-reciprocal displacement coupling.
The latter describes a possible effective linear stability operator, not the exact Jacobian of the minimal particle dynamics.

%--------
\subsection{Mutual contact potential and the passive Hessian}
%-------

At fixed \(\phi\), we take the passive energy to be a sum of one-sided pair potentials,
\begin{equation}
 E_0=\sum_{i<j} V_c(r_{ij};\phi),\qquad
 V_c(r_{ij};\phi)=\frac{k}{a}\,\delta_{ij}^{\,a}\Theta(\delta_{ij}),
 \quad
 \delta_{ij}=1-\frac{r_{ij}}{\sigma_{ij}(\phi)} .
\end{equation}
Here \(r_{ij}=|\bm r_i-\bm r_j|\), \(\sigma_{ij}(\phi)\) is the effective contact distance, \(k\) is an energy scale, and \(\Theta\) is the Heaviside function. 
The choices \(a=2\) and \(a=5/2\) give harmonic contacts and the three-dimensional Hertz law, respectively.

For a fixed contact network, the microscopic Hessian is
\begin{equation}
 \left(\mathcal H_{\rm mic}\right)_{i\alpha,j\beta}
 =\frac{\partial^2 E_0}{\partial r_{i,\alpha}\partial r_{j,\beta}}.
 \label{eq:passive-hessian-mic}
\end{equation}
It is real symmetric and positive semidefinite at a passive local energy minimum. 
Exact symmetry zero modes, such as translations under periodic boundary conditions, are removed when discussing a static inverse. 
The Hessian contains both longitudinal contact stiffness and geometric terms proportional to the contact forces \cite{Wyart2005,Manning2015}. 
The unshifted Wishart matrix \(\mathcal H_0\) introduced in Sec.~\ref{sec:RMT_tumor} is a surrogate for the passive spectrum on the retained displacement subspace; it is not an elementwise model of Eq.~\eqref{eq:passive-hessian-mic}. 
A prestress shift is discussed separately in Appendix~\ref{SubSec:Appl_RMT}.

%-------------
\subsection{Self-propulsion and the displacement Jacobian}
%------------

For a minimal model with substrate drag \(\zeta>0\), the translational dynamics is
\begin{equation}
 \zeta\dot{\bm r}_i=-\frac{\partial E_0}{\partial\bm r_i}
                      +F_a\bm n_i,
 \label{Eq:motion}
\end{equation}
where \(F_a\) is the propulsion force magnitude and \(\bm n_i\) is a unit director. 
For example, torque-free rotational diffusion can be written in Stratonovich form with Stratonovich product \(\circ\) as
\begin{equation}
 d\bm n_i=\sqrt{2D_{\rm r}}\,
 \bigl(\mathbb I_D-\bm n_i\bm n_i^T\bigr)\circ d\bm W_i,
 \label{eq:director}
\end{equation}
with rotational diffusivity \(D_{\rm r}\) and independent standard \(D\)-dimensional Wiener processes \(\bm W_i\), where \(\mathbb{I}_D\) is the \(D\)-dimensional unit matrix. 
In two dimensions, the director angle obeys a scalar diffusion equation.  
In a run-and-tumble model, the director instead reorients at discrete random events \cite{Tailleur2008,Solon2015,CatesTailleur2013}.
Both examples have director dynamics independent of displacements. At \(F_a=0\) the director drops out of the translational equation, regardless of its dynamics.

Consider displacement perturbations about a reference configuration or trajectory, with a fixed contact network for the linearization.
At fixed directors, the direct derivative of \(F_a\bm n_i\) with respect to any particle position is zero. 
Allowing the directors to fluctuate adds \(F_a\,\delta\bm n_i\) to the linearized force. 
In the torque-free model, this is a forcing term, not displacement feedback. 
Activity can alter the reference trajectory and hence the contact Hessian evaluated on it, but Eqs.~\eqref{Eq:motion} and \eqref{eq:director} do not generate an additive non-symmetric displacement Jacobian. Breaking detailed balance does not by itself imply such a term.

Non-reciprocal displacement couplings require additional physics, for example contact-dependent reorientation or an explicitly non-reciprocal interaction \cite{Fruchart21}. 
For an instantaneous effective active force \(\bm F_{\rm eff}(\bm r)\), the relevant expansion would be
\begin{equation}
 \bm F_{\rm eff}(\bm r_*+\delta\bm r)
 -\bm F_{\rm eff}(\bm r_*)
 =\alpha_0\mathcal A\,\delta\bm r
   +O(\|\delta\bm r\|^2).
 \label{active_force}
\end{equation}
Here \(\bm r\) collects all retained displacement coordinates, \(\bm r_*\) is the reference configuration, and \(\alpha_0\mathcal A\) is the effective force derivative on that subspace. 
The reference force is already included in the reference motion and does not enter this derivative. If displacement and orientation are coupled, eliminating the directors generally produces a frequency-dependent response; a time-local coupling as in Eq.~\eqref{active_force} is an additional modeling assumption.

In Sec.~\ref{sec:RMT_tumor} we adopt \(\mathcal M=-\mathcal H_0+\alpha_0\mathcal A\) with an independent real Ginibre matrix \(\mathcal A\). 
In units of the passive spectral scale, \(\alpha_0\) measures the strength of the modeled random coupling. 
It is not identified with the microscopic force \(F_a\) or with the tumbling rate. The Hermitized calculation in Appendix~\ref{app:corrected-hermitization}, in particular Eqs.~\eqref{eq:c-new-offsupport} and \eqref{eq:c-new-static-threshold}, concerns this ensemble.
In its stable bulk regime, the active coupling amplifies the generalized static response. 
In the unshifted model at \(q=1\), any nonzero \(\alpha_0\) instead places part of the limiting spectrum in the right half-plane, so a finite-matrix inverse there need not describe a dynamically attainable static response.

%%%%%%%%%%%%%%%%%%%%%%%%%%%%%%%%%
\section{Wishart--Ginibre model and static response}
\label{sec:RMT_tumor}
%%%%%%%%%%%%%%%%%%%%%%%%%%%%%

\subsection{Effective operator and ensemble}
%--------------------------

On the retained displacement subspace, we study the linear model
\begin{equation}
 \zeta\,\delta\dot{\bm r}=\mathcal M\,\delta\bm r,
 \label{linear_eq_mod}
\end{equation}
with
\begin{equation}
 \mathcal M=-\mathcal H_0+\alpha_0\mathcal A.
 \label{eq:matrix_decom}
\end{equation}
Here \(\mathcal H_0\) and \(\mathcal A\) are independent random matrices. 
As discussed in Sec.~\ref{sec:active_janus_grains}, this dense ensemble is an effective model, not a derivation of independent matrix elements from the microscopic dynamics.
The drag \(\zeta>0\) sets the time scale and does not affect the sign of the eigenvalue real parts.

Let \(N_c\) be the number of retained displacement coordinates and \(M_c\) the number of constraints in the surrogate. 
We use
\begin{equation}
 \mathcal H_0=\mathcal R\mathcal R^T,\qquad q=\frac{N_c}{M_c}\leq 1,
 \label{eq:wishart_def}
\end{equation}
where \(\mathcal R\) is an \(N_c\times M_c\) matrix of independent
real Gaussian entries with mean zero and variance \(1/M_c\).
In the simplest frictionless contact counting, \(N_c\simeq DN\)
and \(M_c=NZ/2\), so that in the thermodynamic limit
\begin{equation}
 q\simeq\frac{2D}{Z}.
 \label{eq:qZ}
\end{equation}
Thus \(q=1\) represents the isostatic limit of the model. The
passive spectral scale is set to unity. The active matrix
\(\mathcal A\) is real Ginibre:
\begin{equation}
 \langle\mathcal A_{ij}\rangle=0,\qquad
 \langle\mathcal A_{ij}\mathcal A_{k\ell}\rangle
 =\frac{1}{N_c}\delta_{ik}\delta_{j\ell}.
 \label{Ginibre}
\end{equation}
In particular, the symmetric component of \(\mathcal A\) is not
removed. The parameter \(\alpha_0\) controls the strength of
this full random coupling in passive spectral units.

At fixed \(q\), the limiting passive density on
\([\lambda_-,\lambda_+]\) is
\begin{equation}
 \rho_0(\lambda)=
 \frac{\sqrt{(\lambda_+-\lambda)(\lambda-\lambda_-)}}{2\pi q\lambda},
 \qquad \lambda_\pm=(1\pm\sqrt q)^2 .
 \label{rho_0_MP}
\end{equation}
As \(q\uparrow 1\), the lower edge vanishes as
\(\lambda_-\sim(1-q)^2/4\). At \(q=1\),
\begin{equation}
 \rho_0(\lambda)\sim \frac{1}{\pi\sqrt{\lambda}}
 \qquad (\lambda\downarrow0).
 \label{rho_0}
\end{equation}
An optional isotropic prestress shift is discussed in
Appendix~\ref{SubSec:Appl_RMT}; all results below are for the
unshifted ensemble.

\subsection{Hermitization and the stable static response}

For an isotropic static test force with
\(\langle\bm f\bm f^\dagger\rangle_f=\sigma_f^2\mathbb I_{N_c}\),
linear stability ensures that
\(\delta\bm r(t)\to-\mathcal M^{-1}\bm f\) when this force is
added to Eq.~\eqref{linear_eq_mod}. The resulting generalized
mean-square susceptibility is
\begin{equation}
 J=\frac{\langle\|\mathcal M^{-1}\bm f\|^2\rangle_f}
         {N_c\sigma_f^2}
  =\frac{1}{N_c}\operatorname{Tr}
    [(\mathcal M\mathcal M^\dagger)^{-1}]
  =\frac{1}{N_c}\sum_k\sigma_k^{-2},
 \label{eq:compliance_def}
\end{equation}
where \(\sigma_k\) are the singular values of \(\mathcal M\);
see Appendix~\ref{app:compliance_derivation}. This is not a shear
compliance. For a non-normal matrix it cannot in general be
inferred from the magnitudes of its eigenvalues alone. If
\(\mathcal M\) is invertible but unstable, the inverse trace
is still defined, but it is not a long-time response.

For a complex probe \(z\), write \(B=z\mathbb I_{N_c}-\mathcal M\).
Girko's Hermitization \cite{Girko1985} gives a block resolvent
whose diagonal traces determine
\begin{equation}
 J_{\epsilon,N_c}(z)=\frac{1}{N_c}\operatorname{Tr}
  \bigl[(BB^\dagger+\epsilon^2\mathbb I_{N_c})^{-1}\bigr],
 \qquad \epsilon>0 .
 \label{def:G}
\end{equation}
We first take \(N_c,M_c\to\infty\) at fixed \(q\) and \(\epsilon\).
The regulated large-matrix trace \(J_\epsilon(z)\) obeys the
Dyson equation derived in Appendix~\ref{app:corrected-hermitization},
in particular Eq.~\eqref{eq:c-new-selfconsistent}:
\begin{gather}
 K(\eta;z)=\int_{\lambda_-}^{\lambda_+}
       \frac{\rho_0(\lambda)\,d\lambda}{|z+\lambda|^2+\eta^2},
 \notag\\
 \eta=\epsilon[1+\alpha_0^2J_\epsilon(z)],\qquad
 J_\epsilon(z)=\frac{\eta}{\epsilon}K(\eta;z).
 \label{eq:SCBA_MP}
\end{gather}
The self-energy thus renormalizes the imaginary regulator.
It is not an activity-induced positive mass that survives when
\(\epsilon\downarrow0\) on the exterior branch.

Outside the limiting eigenvalue support, where
\(\alpha_0^2K(0;z)<1\), the zero-regulator limit is
\(J(z)=K(0;z)/[1-\alpha_0^2K(0;z)]\), as given in
Eq.~\eqref{eq:c-new-offsupport}. For \(q<1\) and
\(\operatorname{Re}z\geq0\), \(K(0;z)\) is largest at \(z=0\).
The first contact of the limiting support with the imaginary axis
therefore occurs at \(z=0\)
(Appendix~\ref{app:c-stability}). The passive inverse-square
moment at this point is
\begin{equation}
 \mathcal I_0(q)=K(0;0)
 =\int_{\lambda_-}^{\lambda_+}\frac{\rho_0(\lambda)}{\lambda^2}\,d\lambda
 =(1-q)^{-3},
 \label{def:I_0}
\end{equation}
as evaluated in Eq.~\eqref{eq:c-new-mp-m2}. Accordingly, the
bulk stability threshold and static susceptibility are
\begin{equation}
 \alpha_c=(1-q)^{3/2},\qquad
 J=J(0)=\frac{(1-q)^{-3}}{1-(\alpha_0/\alpha_c)^2},
 \quad 0\leq\alpha_0<\alpha_c .
 \label{eq:SCBA_explicit}
\end{equation}
See Eq.~\eqref{eq:c-new-static-threshold}. For fixed \(q<1\),
the weak-activity expansion reads
\begin{equation}
 J=\mathcal I_0(q)+\alpha_0^2\mathcal I_0(q)^2
     +O(\alpha_0^4),
 \label{eq:G_non_jammed}
\end{equation}
with the positive coefficient independently checked in Appendix~\ref{app:c-check}. 
Thus, activity amplifies the susceptibility in the stable bulk, rather than suppressing it.

%%%%%%%%
\subsection{Isostatic limit and finite-size interpretation}
\label{subsec:offcritical_scope}
%%%%%%%%%

At \(q=1\) the passive inverse-square moment diverges, and the
unshifted limiting spectrum enters the right half-plane for any
\(\alpha_0>0\). There is no stable isostatic branch with a finite
static susceptibility at nonzero activity. Inside the eigenvalue
support the regulated squared resolvent diverges as
\(\epsilon\downarrow0\) according to
Eq.~\eqref{eq:c-new-interior}; this is not an activity power law.

These are thermodynamic-first results. At finite \(N_c\) the sign of the largest eigenvalue real part must be checked for each realization. 
The finite-size unregularized inverse trace at positive activity is heavy-tailed; its disorder mean is not the quantity in Eq.~\eqref{eq:SCBA_explicit}. 
A comparison with direct matrix data should distinguish typical values (or regulated traces) from the fraction of dynamically stable realizations.

%%%%%%%%%%%%%%%%%%%%%%%%%%%%%%%
\section{Direct random-matrix tests}
\label{sec:direct_random_matrix_results}
%%%%%%%%%%%%%%%%%%%%%%%%%%%

\begin{figure}[t]
 \centering
 \includegraphics[width=\linewidth]{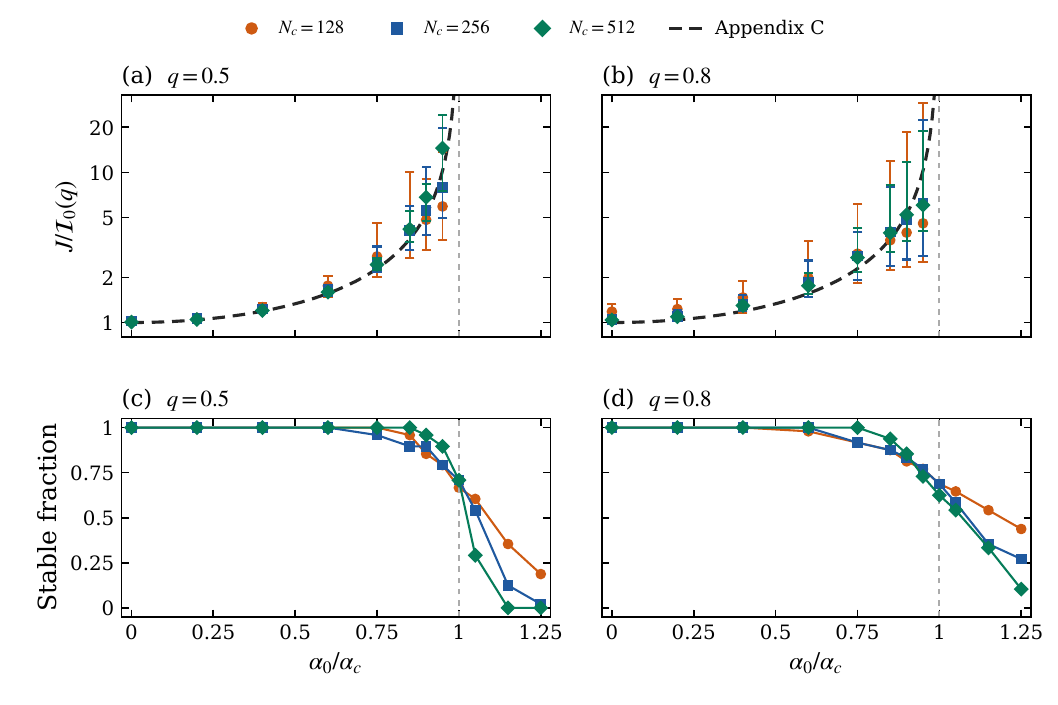}
 \caption{Stable-side susceptibility and finite-size stability of the Wishart--Ginibre ensemble. 
 The columns correspond to \(q=0.5\) and \(q=0.8\). 
 (a,b) The symbols show the median of \(J/\mathcal I_0(q)\) over linearly stable realizations at each \(\alpha_0<\alpha_c\); bars span the 25th--75th percentiles. 
 The dashed curve is the thermodynamic prediction \(1/[1-(\alpha_0/\alpha_c)^2]\), with no fitted parameters. 
 (c,d) The fraction of all realizations at each point with \(\max_k\operatorname{Re}z_k<0\), where \(z_k\) are the eigenvalues of \(\mathcal M\). 
 The vertical line marks the limiting bulk threshold \(\alpha_0=\alpha_c\), not a finite-sample criterion.
 Each pair \((q,N_c)\) uses 48 independent draws; the same matrices are used at all activities within each draw.}
 \label{fig:stable_svd_test}
\end{figure}

%%%%%%%%%%%%%%%%%%%%%%%%
\subsection{Direct test of the stable-side susceptibility}
%%%%%%%%%%%%%%%%%%%%%%%

To test Eq.~\eqref{eq:SCBA_explicit}, we evaluate \(J\) by a singular
value decomposition of
\(\mathcal M=-\mathcal H_0+\alpha_0\mathcal A\) and determine stability
independently from the complex eigenvalues of the same matrix. We
draw \(\mathcal H_0=\mathcal R\mathcal R^T\) and \(\mathcal A\)
independently, using the Gaussian variances in
Eqs.~\eqref{eq:wishart_def} and \eqref{Ginibre}. We use
\(N_c=128,256,512\) and \(M_c=N_c/q\) at \(q=0.5\) and \(0.8\).
The normalization is the thermodynamic passive value
\(\mathcal I_0(q)=(1-q)^{-3}\), not a fitted or sample-dependent
baseline. We use quantiles rather than an unregularized disorder
mean for the finite-size inverse trace
(Appendix~\ref{app:corrected-hermitization}).

Figures~\ref{fig:stable_svd_test}(a,b) show that the typical response increases with activity within the stable regime, as predicted by
Eq.~\eqref{eq:c-new-static-threshold}. 
For example, at \(N_c=512\) and \(\alpha_0/\alpha_c=0.60\), all draws are stable and the median normalized responses are \(1.60\) and \(1.76\) for \(q=0.5\) and \(0.8\), respectively, compared with the limiting value \(1.5625\).
Near the threshold, the response becomes more dispersed, and Figs.~\ref{fig:stable_svd_test}(c,d) show that stable and unstable realizations coexist in the finite-size samples. 
The medians in this region are conditional on stability and are not a numerical measurement of the thermodynamic divergence.

%%%%%%%%%%% 
\subsection{Smallest-singular-value contribution near marginality}
\label{subsec:smallest_singular_share}
%%%%%%%%%%%%

Near the isostatic limit, it is useful to ask how much of the
finite-matrix inverse trace comes from a single singular value.
For the singular values \(\sigma_k\) of \(\mathcal M\), define
\begin{equation}
 J_{\min}=\frac{1}{N_c\sigma_{\min}^{2}},\qquad
 R_{\min}=\frac{J_{\min}}{J}
 =\left[\sum_{k=1}^{N_c}
        \left(\frac{\sigma_{\min}}{\sigma_k}\right)^2\right]^{-1}.
 \label{eq:smallest_singular_share}
\end{equation}
A value near unity means that the smallest singular value alone
carries most of \(J\). This is a finite-size diagnostic, not a test of
an activity exponent or a criterion for linear stability.

\ifdefined\IVBUseArchivedSummaries
\begin{figure}[t]
 \centering
 \includegraphics[width=0.79\linewidth]{Fig2_from_summaries.pdf}
 \caption{Smallest-singular-value share \(R_{\min}=J_{\min}/J\)
 at \(q=1000/1001\) for \(N_c=1000,2000,4000\).
 Symbols are medians and bars span the 25th--75th percentiles
 calculated across the archived realizations. The horizontal line
 is the single-mode limit \(R_{\min}=1\).
 Stability was not resolved in the archived summaries.}
 \label{fig:alpha_dep}
\end{figure}
In Fig.~\ref{fig:alpha_dep}, the share is often large, indicating
that the inverse trace is sensitive to the bottom of the singular
spectrum. Because these archived summaries do not provide the
spectral abscissa of the individual matrices, they are interpreted
as statistics of the inverse trace, not as measurements of the
steady susceptibility.
\else
\begin{figure}[t]
 \centering
 \includegraphics[width=0.79\linewidth]{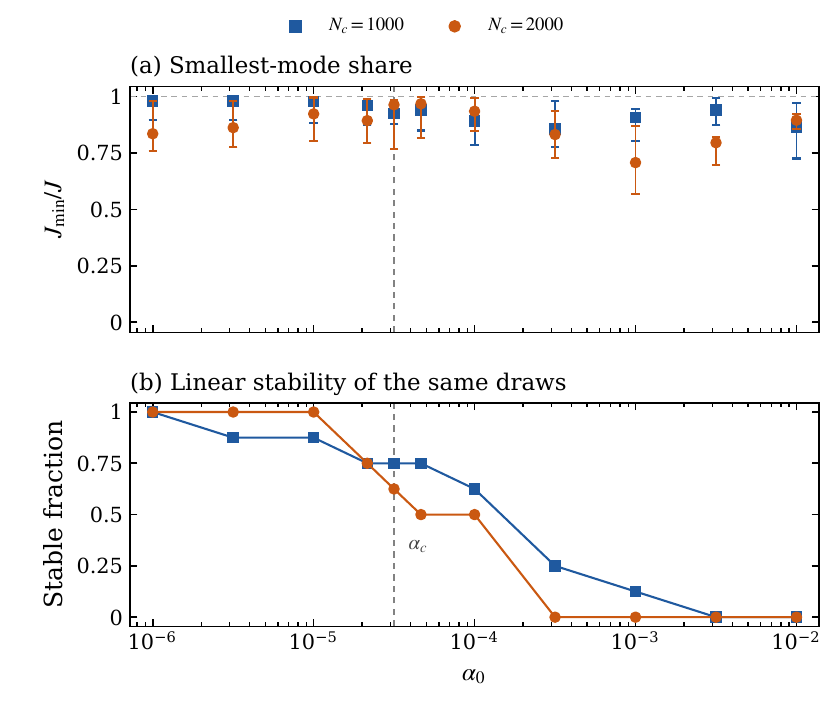}
 \caption{Smallest-singular-value share and linear stability in independent near-marginal simulations at \(q=1000/1001\).
 (a) Median \(R_{\min}=J_{\min}/J\) and interquartile range across all realizations at each activity. The horizontal line marks \(R_{\min}=1\). 
 (b) Fraction of the same realizations with spectral abscissa \(\max_k\operatorname{Re}z_k<0\).
 The vertical line in both panels is the thermodynamic bulk threshold \(\alpha_c=(1-q)^{3/2}\simeq3.16\times10^{-5}\), not a stability test for individual matrices. We use eight independent draws at each of \(N_c=1000\) and \(2000\).
 Within a draw, the passive and active matrices are held fixed across activities. 
 For the same passive Wishart realizations at \(\alpha_0=0\), the median \(R_{\min}^{(0)}\) is \(0.982\) (\(0.898\)--\(0.997\) IQR) for \(N_c=1000\) and \(0.830\) (\(0.743\)--\(0.979\) IQR) for \(N_c=2000\).}
 \label{fig:alpha_dep}
\end{figure}

Figure~\ref{fig:alpha_dep}(a) shows that the smallest singular
value can carry much of the inverse trace on both sides of the
thermodynamic threshold. Panel (b) shows why this observation
must be distinguished from an enhanced steady response:
finite matrices that have a large smallest-mode share can still
be linearly unstable. We retain all realizations in panel (a)
to characterize the matrix statistic and use the independently
computed spectral abscissa to classify stability. For example, at
\(\alpha_0=\alpha_c\) only six of the eight \(N_c=1000\) draws and
five of the eight \(N_c=2000\) draws are stable. These counts
illustrate the finite-sample spread around the bulk threshold,
rather than locating a new threshold.

To determine how much of this single-mode dominance is already present without activity, we also evaluate \(R_{\min}\) at
\(\alpha_0=0\) for the same Wishart realizations used in Fig.~\ref{fig:alpha_dep}. 
The passive median values are \(R_{\min}^{(0)}=0.982\) and \(0.830\) for \(N_c=1000\) and \(2000\), respectively, with interquartile ranges \(0.898\)--\(0.997\) and \(0.743\)--\(0.979\). 
These values are already large and are close to those at the smallest nonzero activity shown in Fig.~\ref{fig:alpha_dep}. 
The passive lower edge is of order \((1-q)^2\). 
The Marchenko--Pastur weight on that scale is of order
\(1-q\), so these sizes contain only of order \(N_c(1-q)\) passive
modes on the cutoff scale. 
The direct passive comparison
therefore confirms that the strong smallest-mode weight at weak activity
is largely inherited from the near-isostatic passive spectrum and should
not be attributed to activity-induced non-normality alone.
\fi

Thus Fig.~\ref{fig:alpha_dep} complements Fig.~\ref{fig:stable_svd_test}:
the latter tests the stable-side thermodynamic prediction away from
the isostatic limit, whereas the former documents how a single
singular value can dominate the finite-size inverse trace near
isostaticity. Neither figure implies a finite steady susceptibility
at \(q=1\) and nonzero activity; that regime is unstable in the
unshifted large-matrix model
(Appendix~\ref{app:c-stability}).

\ifdefined\IVBIncludeStabilityMap
\begin{figure}[t]
 \centering
 \includegraphics[width=0.79\linewidth]{Fig3_stability_map.pdf}
 \caption{Stability diagram of the unshifted
 Wishart--Ginibre model. (a) At \(N_c=512\), symbols show the
 fraction of linearly stable realizations at each pair
 \((1-q,\alpha_0)\); \(q=N_c/M_c\) uses the actual integer dimensions.
 The dashed line is the analytic bulk boundary
 \(\alpha_c=(1-q)^{3/2}\), with no fit to the numerical points.
 (b) Finite-size stability fractions on the \(q=1\) slice.
 Each symbol uses 16 independent draws.}
 \label{fig:stability_map}
\end{figure}
Figure~\ref{fig:stability_map}(a) illustrates how the width of
the stable bulk region collapses as \(q\) approaches unity.
At \(q=1\), a finite matrix may remain stable at small activity,
as seen in panel (b); this is not a nonzero thermodynamic
threshold. The dashed boundary represents the limiting spectrum,
whereas the symbols represent finite-size ensemble fractions.
\fi

%%%%%%%%%%%%%%%%%%%%%%%%%%%%%
\section{Discussion}\label{Sec:Discussion}
%%%%%%%%%%%%%%%%%%%%%%

\subsection{Bulk amplification and finite-size stability}
\label{subsec:finite_size_extreme}
%%%%%%%%%%%%%%%%%%%%

In the unshifted model, the same passive inverse-square moment sets both the strength of the zero-frequency response and the instability threshold. Increasing activity at fixed \(q<1\) raises \(J\) on the stable side [Eq.~\eqref{eq:c-new-static-threshold}]. 
At the bulk threshold, the limiting spectrum reaches the imaginary axis at zero.
This is not a positive elastic shift of the passive modes.
In particular, a finite value of the inverse trace for a specific sample at \(q=1\) and nonzero activity does not establish a stable thermodynamic response.

The large-system calculation takes \(N_c\to\infty\) at fixed positive regulator, and only then sends the regulator to zero (Appendix~\ref{app:corrected-hermitization}). 
This order cannot be replaced by averaging the unregularized static trace of finite matrices. 
For positive activity, the real Gaussian ensemble permits matrices arbitrarily close to a simple rank deficiency, and the unregularized finite-size mean is infinite.
The medians in Fig.~\ref{fig:stable_svd_test} are instead taken over stable realizations, with the stability fraction shown independently. 
Their increase with activity is consistent with the bulk prediction away from the threshold, but their behavior near the threshold does not measure the large-system divergence.

Closer to isostaticity, Fig.~\ref{fig:alpha_dep} shows that much of the finite-size trace can come from a single singular value.
This can happen even at weak activity, when the passive soft modes already carry a large share. 
A small singular value is not by itself evidence of non-normality or a test of stability.
For example, for an invertible matrix the Schur decomposition of \(\mathcal M^{-1}\) gives
\[
 J=\frac{\|\mathcal M^{-1}\|_{\rm F}^{2}}{N_c}
 \ \geq\ \frac{1}{N_c}\sum_k |z_k|^{-2},
\]
with equality for a normal matrix, where \(z_k\) are the eigenvalues of \(\mathcal M\) and \(\|\cdot\|_{\rm F}\) is the Frobenius norm.
The smallest-mode share and the difference from an eigenvalue-only proxy therefore answer different questions.

%%%%%%%%%%%%%%%%
\subsection{Relation to active materials and possible tests}
\label{sec:discussion_tests}
%%%%%%%%%%%%%

Prior work treats response and stability for additive matrix deformations \cite{AFM2015,HoZhong2023}. 
For the Wishart passive spectrum, we show that the stable activity window shrinks with width \(\alpha_c=(1-q)^{3/2}\) as \(q\to1^-\), and that, at finite size, dominance by the smallest singular value must be assessed separately from the existence of a steady response. 
With the frictionless contact-counting identification \(q\simeq2D/Z\), the threshold becomes Eq.~\eqref{def:alpha_c} near isostaticity.  
The \(3/2\) collapse of the stability window with excess coordination is therefore a testable prediction of the present surrogate, provided that the strength of the additional displacement feedback is determined independently rather than inferred from the observed instability threshold.

To make the connection between the random-matrix parameter \(\alpha_0\) and a microscopic model more explicit, consider a fixed mechanically stable configuration of the microscopic system.  
Let \(\mathcal M_{\rm mic}\) denote the Jacobian governing infinitesimal displacements about this configuration.  
We separate it into the passive elastic contribution and the remaining displacement-dependent feedback as
\begin{equation}
 \mathcal M_{\rm mic}
 =
 -\mathcal H_{\rm passive}
 +\mathcal K_{\rm fb}.
\end{equation}
Here \(\mathcal H_{\rm passive}\) is the Hessian of the passive interaction energy evaluated at the same configuration, while \(\mathcal K_{\rm fb}\) is defined as the remainder
\begin{equation}
 \mathcal K_{\rm fb}
 :=
 \mathcal M_{\rm mic}+\mathcal H_{\rm passive}.
\end{equation}
Thus \(\mathcal K_{\rm fb}\) contains those contributions to the linearized force that depend on the displacement but are not contained in the passive Hessian.  
In particular, a position-independent self-propulsion force at fixed orientation does not contribute to \(\mathcal K_{\rm fb}\), as discussed in Sec.~\ref{sec:active_janus_grains}.  
Depending on the microscopic mechanism, \(\mathcal K_{\rm fb}\) may contain both reciprocal and non-reciprocal components and need not be antisymmetric.

Equation~\eqref{eq:matrix_decom} is the random-matrix surrogate of this microscopic decomposition: \(\mathcal H_0\) replaces \(\mathcal H_{\rm passive}\) by a Wishart matrix, while \(\alpha_0\mathcal A\) models \(\mathcal K_{\rm fb}\) by an independent full real Ginibre matrix.  
This correspondence is statistical rather than an element-by-element identification.

It also suggests an operational way of estimating the strength of the feedback independently of the instability threshold.  
After expressing \(\mathcal K_{\rm fb}\) in the same passive spectral units used for \(\mathcal H_{\rm passive}\), define
\begin{equation}
 \alpha_{\rm eff}^{2}
 :=
 \frac{1}{N_c}
 \operatorname{Tr}
 \left(
 \mathcal K_{\rm fb}\mathcal K_{\rm fb}^{T}
 \right).
\end{equation}
For the surrogate
\(\mathcal K_{\rm fb}=\alpha_0\mathcal A\), with
\(\langle \mathcal A_{ij}^{2}\rangle=1/N_c\), one has
\(\langle\alpha_{\rm eff}^{2}\rangle=\alpha_0^{2}\).
Thus \(\alpha_{\rm eff}\) can be measured from the microscopic linearized operator without using the observed stability threshold.

This construction does not assume that the microscopic \(\mathcal K_{\rm fb}\) has independent Ginibre entries.  
It may be sparse, spatially structured, correlated with the passive network, or frequency dependent.  
The predicted threshold should therefore be regarded as a prediction of the Wishart--Ginibre surrogate whose robustness to these microscopic structures remains to be tested.

Neither independent Ginibre couplings nor the numerical threshold should be assumed to describe a particular tissue or active packing. Real contact operators are spatially structured, and their active couplings can be correlated with the passive network. 
Appendices~\ref{app:cellular_automaton} and \ref{app:alpha_derivation} illustrate how additional sensing laws can produce feedback without specifying its statistics. 
Prestress can shift the passive edge (Appendix~\ref{SubSec:Appl_RMT}); contact rearrangements or slow orientation modes can invalidate a time-independent displacement operator altogether.

The fixed-network surrogate also leaves out important aspects of marginal stability in particle packings. 
Although the Wishart spectrum represents a population of soft modes approaching isostaticity, the model contains neither the forces carried by individual contacts nor the gaps between pairs of particles that are not in contact. 
It therefore cannot predict the critical distributions of weak contact forces and small gaps, the opening of weak contacts, the formation of new contacts, or how these statistics are modified by activity. Such effects become essential once the perturbation or the active dynamics changes the contact network.
The present results should consequently be interpreted as predictions for the linear response and stability of a fixed-network surrogate before contact rearrangements occur, rather than as a complete theory of marginal stability in an active jammed packing.

The most direct test would apply several weak, independent static force patterns to the same mechanically stable configuration and measure the mean-square incremental displacement. 
An isotropic set of forces probes the trace in Eq.~\eqref{eq:compliance_def}; localized or shear loading weights the response differently (Appendix~\ref{app:compliance_derivation}). 
Displacement fluctuations in an unforced active material are not, in general, a measure of this static trace: even for linear dynamics they depend on the time spectrum and spatial covariance of the driving. Simulations that measure both the displacement Jacobian and the response to controlled forces could test which features of the model survive spatial correlations, prestress, and changes in the contact network.

\section{Conclusion}\label{Sec:Conclusion}

For an unshifted Wishart operator with an independent full
real Ginibre perturbation, the squared static response
increases with activity as long as the limiting spectrum
is stable. At fixed \(q<1\) it is
\(J=(1-q)^{-3}/[1-(\alpha_0/\alpha_c)^2]\), with
\(\alpha_c=(1-q)^{3/2}\). At \(q=1\) the bulk spectrum
is unstable for every \(\alpha_0>0\); activity does not
regularize a steady isostatic response.

Direct matrix calculations show amplification in the stable regime and distinguish it from finite-size effects near marginality. 
In particular, the smallest singular value can dominate the inverse trace even when the matrix is unstable.
Reporting response statistics together with the fraction of stable samples is therefore essential to interpreting the near-marginal data. These predictions concern an isotropically forced effective linear network, not a conventional shear modulus or a complete microscopic theory of marginal stability in an active jammed material. 
In particular, the fixed-network model does not describe weak-force and small-gap statistics or contact rearrangements.

\begin{acknowledgments}
This paper is inspired by Suravi Pal's presentation at the mini-workshop on ``the rheology of dense materials" (YITP-X-25-09).
The author appreciates stimulating discussions with Suravi Pal and Takeshi Kawasaki.
He also thanks G. Szamel for his critical reading.
This work was partially supported by the JSPS KAKENHI Grant No. JP26K06960.
This paper is dedicated to Professor Hajime Tanaka to celebrate his scientific achievements.
\end{acknowledgments}

\section*{AUTHOR DECLARATIONS}

\subsection*{Conflict of Interest}
% AUTHOR: Confirm your disclosure before submission. If there are no
% conflicts, replace the placeholder below with:
% The author has no conflicts to disclose.
The author has no conflicts of interest to disclose.

\subsection*{Author Contributions}
% AUTHOR: Suggested CRediT roles; retain only those that match your work.
\textbf{Hisao Hayakawa}: Conceptualization; Formal analysis; Methodology;
Writing -- original draft; Writing -- review \& editing.

\section*{DATA AVAILABILITY}

The realization-level data and scripts used for the numerical figures are available from the corresponding author upon reasonable request.

\newpage

%%%%%%%%%%
\appendix
%%%%%%%%%%%%%%%%%%%%%%%%%%%%
\section{Displacement feedback from internal degrees of freedom}
\label{app:cellular_automaton}

\subsection{Active forcing is not the same as active feedback}
%%%%%%%%%%%%%%%%%%%%%%%%%%

For the torque-free model in Sec.~\ref{sec:active_janus_grains}, the
propulsion force can do work even though it has no direct
position derivative. In the absence of probe forces, and for a
time-independent passive contact energy, Eq.~\eqref{Eq:motion} gives
\begin{equation}
 \frac{dE_0}{dt}
 =-\zeta\sum_i|\dot{\bm r}_i|^2
   +F_a\sum_i\bm n_i\cdot\dot{\bm r}_i .
 \label{eq:app-active-work}
\end{equation}
Neither a nonzero active power nor a violation of detailed balance
establishes a non-symmetric displacement Jacobian. In particular,
rotational diffusion or independent run-and-tumble reorientation
\cite{Tailleur2008,Solon2015} does not sense a displacement perturbation
in the model considered here. Activity may nevertheless change the
trajectory and the sequence of contact networks. Linearization along
such a trajectory generally gives time-dependent coefficients, not
the fixed random matrix in Eq.~\eqref{eq:matrix_decom}.

\subsection{Eliminating an internal variable}

Consider a displacement-sensitive internal variable instead, and linearize about a fixed reference state on the retained displacement subspace. 
Let \(\bm u\) be the displacement perturbations and \(\bm p\) be perturbations of the internal variables (for example, contact-dependent orientation or active tensions). 
If the extra feedback is mediated by these variables, the linear equations are
\begin{equation}
 \zeta\dot{\bm u}=-H\bm u+B\bm p+\bm f,\qquad
 \dot{\bm p}=C\bm u-G\bm p+\bm h ,
 \label{eq:app-feedback-block}
\end{equation}
where \(H=H^T\) is the potential Hessian at fixed internal state,
\(B\) maps internal variables to forces, \(C\) describes their response to displacement, \(G\) is their relaxation operator, and \(\bm f,\bm h\) are test sources. 
These block equations assume that
the retained first-order displacement dynamics have scalar drag.

For a Laplace variable \(s\), neglecting initial-condition terms or treating them as additional sources, eliminating \(\bm p\) gives the exact reduced equation
\begin{align}
 \bigl[s\zeta I+H-B(sI+G)^{-1}C\bigr]\bm u(s)
 &=\bm f(s)+B(sI+G)^{-1}\bm h(s),\\
 \mathcal M_{\rm eff}(0)&=-H+BG^{-1}C .
 \label{eq:app-feedback-schur}
\end{align}
The second line is the static Schur complement for invertible
\(G\); its interpretation as a long-time static response requires
stability of the full coupled dynamics. In general it is not the
operator governing displacement relaxation. For example, for
\(|s|\,\|G^{-1}\|\ll1\), the inverse response on the left has
the expansion
\begin{equation}
 H-BG^{-1}C+s\bigl[\zeta I+BG^{-2}C\bigr]+O(s^2).
 \label{eq:app-feedback-memory}
\end{equation}
Using only \(\mathcal M_{\rm eff}(0)\) and the original drag for
slow dynamics also requires the neglected frequency dependence
to be small on the relevant modes; when it is not, the internal
dynamics or the induced memory must be retained. Stability of
\(G\) alone does not ensure stability of the coupled system.

The feedback vanishes for \(C=0\), as in the torque-free
linearization. It is symmetric in simple reciprocal examples
(e.g., \(G=G^T\), \(C=B^T\) in compatible units), but symmetry
need not survive when the force and sensing channels are
different. For example, one scalar internal variable \(p\) may
sense the displacement along a contact gradient \(\bm t\) and
produce forces along a distinct gradient \(\bm b\):
\begin{equation}
 \zeta\dot{\bm u}=-H\bm u+\beta\bm b p,\qquad
 \dot p=-\gamma_p p+\kappa_p\bm t^T\bm u,\qquad \gamma_p>0 .
 \label{eq:app-contact-feedback}
\end{equation}
The static correction is
\((\beta\kappa_p/\gamma_p)\bm b\bm t^T\). If \(\bm b\) and
\(\bm t\) are linearly independent on the retained subspace,
this correction is non-symmetric for nonzero \(\beta\kappa_p\).
For example, two distinct edge extensions of a three-particle
configuration can supply these gradients. Each internal tension
may produce equal and opposite forces on its own edge; non-reciprocal
\emph{cross}-responses do not require a violation of pairwise action
and reaction. This example introduces a sensing law as an additional
assumption; it is not a consequence of isotropic self-propulsion.

Such microscopic couplings can be sparse, constrained by geometry,
correlated with \(H\), or strongly frequency dependent. Even a
dense representation in the normal-mode basis need not have
independent entries. The independent Ginibre matrix of
Eq.~\eqref{Ginibre} is a statistical model of a full non-symmetric
perturbation, not a consequence of the block elimination. No
relation between its amplitude \(\alpha_0\) and a microscopic
force or persistence time follows from this construction.

\section{Vertex-model illustration: curvature and tension feedback}
\label{app:alpha_derivation}

\subsection{Curvature at fixed material state}

In a confluent vertex model with fixed cell adjacency, an illustrative
mechanical energy is \cite{Farhadifar2007,Bi2015,Bi2016}
\begin{equation}
 E=\sum_i\left[
 \frac{K_A}{2}(A_i-A_i^*)^2+
 \frac{K_P}{2}(P_i-P_i^*)^2\right]+\sum_e\Lambda_e\ell_e.
 \label{eq:vertex_energy}
\end{equation}
Here \(A_i,P_i\) are areas and perimeters, \(A_i^*,P_i^*\)
are their target values, \(\ell_e\) is the length of edge \(e\),
and \(K_A,K_P,\Lambda_e\) are prescribed material parameters.
For an amplitude \(u\) along a specified vertex displacement
direction, the stiffness along that direction is
\begin{align}
 k_{\rm eff}
 &=\left.\frac{d^2E}{du^2}\right|_{u=0}\nonumber\\
 &=\sum_i K_A\bigl[(A_i')^2+(A_i-A_i^*)A_i''\bigr]
  +\sum_i K_P\bigl[(P_i')^2+(P_i-P_i^*)P_i''\bigr]
  +\sum_e\Lambda_e\ell_e'' ,
 \label{eq:app-vertex-curvature}
\end{align}
where primes denote \(u\) derivatives at the reference state,
with all material parameters held fixed. The terms containing
second derivatives of the geometry have no universal sign
and may soften or stiffen a mode. An imposed change in a target
or material parameter can therefore change the stiffness, but
the fixed-parameter Hessian remains symmetric, even if its
curvature becomes negative. This calculation neither introduces
a non-reciprocal operator nor determines \(\alpha_0\).

\subsection{A conditional example of cross-edge feedback}

A non-symmetric correction is possible if the tension variable on
one edge responds dynamically to deformation of another. For
illustration, let \(p_e\) be a perturbation of the tension on edge
\(e\), so that its force to linear order is
\(-p_e\bm b_e\), where \(\bm b_e=\nabla_{\bm r}\ell_e\) at the
reference state. Suppose the additional constitutive equation is
\begin{equation}
 \tau_e\dot p_e=-p_e+\chi_{ef}\bm b_f^T\bm u,\qquad \tau_e>0 .
 \label{eq:app-vertex-tension}
\end{equation}
This is an assumed feedback law; it does not follow from
Eq.~\eqref{eq:vertex_energy}. Its static contribution to the
displacement force derivative is
\(-\chi_{ef}\bm b_e\bm b_f^T\), with dynamical factor
\(1/(1+s\tau_e)\). The single-edge case \(e=f\) is symmetric.
For two distinct edges with independent gradients, the
correction can be non-symmetric if the reverse feedback is
absent or differs. The sign and size of this correction are
material- and state-dependent. Neither this example nor a
fixed-topology curvature calculation describes cellular
rearrangements or yields the covariance of the Ginibre ensemble.

%%%%%%%%%%%%%%%%%%%%%%%%%%%%%%%%%
\section{Hermitized Dyson equation for the Wishart--Ginibre model}
\label{app:corrected-hermitization}
%%%%%%%%%%%%%%%%%%%%%%%%%%%%%%%

We derive the squared resolvent for the full real Ginibre coupling, with no
assumption that the activity acts as a positive elastic shift. Let
\(\mathcal M=-\mathcal H_0+\alpha_0\mathcal A\), \(N=N_c\), and
\(\operatorname{tr}_N=N^{-1}\operatorname{Tr}\). The independent real entries
of \(\mathcal A\) are centered Gaussians with variance \(1/N\).
We condition on the symmetric passive matrix \(\mathcal H_0\) and use its
orthogonal eigenbasis; the real Ginibre law is invariant under this change
of basis. Its passive empirical spectral measure is assumed to converge to
\(\rho_0(\lambda)\,d\lambda\). In the Wishart model this is the
Marchenko--Pastur law with the normalization of the main text.

%%%%%%%%%%%%%%%
\subsection{Block resolvent and the regulator}
\label{app:c-block-resolvent}
%%%%%%%%%%%%%%%%

For a complex probe \(z\), set \(B=zI-\mathcal M\), \(B_0=zI+\mathcal H_0\)
and
\begin{equation}
 \mathbb H=\begin{pmatrix}0&B\\B^\dagger&0\end{pmatrix}
 =\begin{pmatrix}0&B_0\\B_0^\dagger&0\end{pmatrix}
  -\alpha_0\mathbb W,
 \qquad
 \mathbb W=\begin{pmatrix}0&\mathcal A\\\mathcal A^T&0\end{pmatrix}.
 \label{eq:c-new-hermitization}
\end{equation}
For \(\epsilon>0\), let
\(\mathbb G_{\epsilon,N}=(i\epsilon I_{2N}-\mathbb H)^{-1}\), and denote the conditional expectation of the sample trace
in Eq.~\eqref{def:G} by
\begin{equation}
 \overline J_{\epsilon,N}(z)=
 \operatorname{tr}_N\mathbb E_{\mathcal A}
       (BB^\dagger+\epsilon^2 I)^{-1}.
 \label{eq:c-new-regulated-finite}
\end{equation}
Block inversion gives
\begin{align}
 [\mathbb G_{\epsilon,N}]_{11}
   &=-i\epsilon(BB^\dagger+\epsilon^2I)^{-1},\qquad
 [\mathbb G_{\epsilon,N}]_{22}
   =-i\epsilon(B^\dagger B+\epsilon^2I)^{-1}.
\end{align}
The two diagonal block traces coincide, sample by sample, since \(B\)
is square. In particular, setting
\(g_{jj,N}=\operatorname{tr}_N\mathbb E_{\mathcal A}
 [\mathbb G_{\epsilon,N}]_{jj}\), one has
\begin{equation}
 g_{11,N}=g_{22,N}=-i\epsilon \overline J_{\epsilon,N}.
 \label{eq:c-new-trace-identity}
\end{equation}
Thus the bare regulator multiplies \(\overline J_{\epsilon,N}\) in the trace that
enters the self-energy.

%%%%%%%%%%%%
\subsection{Leading large-system Dyson equation}
\label{app:c-dyson}
%%%%%%%%%%%%

The exact covariance map of \(\mathbb W\), acting on a deterministic
block matrix \(X\), is
\begin{equation}
 \mathcal S_N[X]=\mathbb E_{\mathcal A}(\mathbb W X\mathbb W)
  =\begin{pmatrix}
   (\operatorname{tr}_N X_{22})I & X_{21}^T/N\\[2pt]
   X_{12}^T/N & (\operatorname{tr}_N X_{11})I
  \end{pmatrix}.
 \label{eq:c-new-covariance}
\end{equation}
For example, the upper diagonal block contracts \(\mathcal A X_{22}
\mathcal A^T\), and the upper off-diagonal block contracts
\(\mathcal A X_{21}\mathcal A\). The latter is specific to real Gaussian
entries and is subleading at fixed \(\epsilon\): a resolvent has norm
at most \(1/\epsilon\). Equation~\eqref{eq:c-new-covariance} is a
finite-\(N\) covariance identity, not a closed equation for the exact
finite-\(N\) averaged resolvent.

With \(\mathbb H_0=\mathbb H+\alpha_0\mathbb W\), Gaussian integration by
parts gives the exact identity
\begin{equation}
 (i\epsilon I-\mathbb H_0)\mathbb E\mathbb G_{\epsilon,N}
 =I+\alpha_0^2\mathbb E\bigl[
 \mathcal S_N[\mathbb G_{\epsilon,N}]\mathbb G_{\epsilon,N}\bigr].
 \label{eq:c-new-ibp}
\end{equation}
At fixed \(\epsilon>0\), the bounded normalized
Gaussian resolvent traces concentrate, so their products with resolvent
blocks factorize to leading order. This gives the matrix Dyson
equation with self-energy \(\alpha_0^2\mathcal S_N[\mathbb G]\), where
\(\mathbb G\) denotes the self-consistent resolvent. Equivalently this is
the deformed Ginibre large-system resolvent construction of
Ref.~\cite{AFM2015} for an orthogonally diagonalizable passive matrix.
We first send \(N\to\infty\) with \(\epsilon>0\) fixed. Keeping the
trace terms in Eq.~\eqref{eq:c-new-covariance}, the two diagonal
self-energies are then the same,
\(\Sigma_{11}=\Sigma_{22}=-i\alpha_0^2\epsilon J_\epsilon I\),
with \(J_\epsilon=\lim_{N\to\infty}\overline J_{\epsilon,N}\).
The diagonal entry of the inverse Dyson matrix is therefore
\(i\eta I\), with \(\eta=\epsilon(1+\alpha_0^2J_\epsilon)\).
The upper diagonal block of its inverse is
\(-i\eta(B_0B_0^\dagger+\eta^2 I)^{-1}\). Taking its normalized
trace and using Eq.~\eqref{eq:c-new-trace-identity} gives
\begin{align}
 K(\eta;z)&=\int
  \frac{\rho_0(\lambda)\,d\lambda}{|z+\lambda|^2+\eta^2},
 \label{eq:c-new-kernel}\\
 J_\epsilon(z)
   &=\frac{\eta}{\epsilon}K(\eta;z)
    =(1+\alpha_0^2J_\epsilon)K\bigl(\epsilon(1+\alpha_0^2J_\epsilon);z\bigr),
 \label{eq:c-new-selfconsistent}\\
 \epsilon&=\eta\,[1-\alpha_0^2K(\eta;z)].
 \label{eq:c-new-eta}
\end{align}
The physical solution has \(\eta>0\) with the bracket positive. On that
branch the right-hand side of Eq.~\eqref{eq:c-new-eta} is strictly
increasing in \(\eta\), since \(K\) decreases. The activity renormalizes
the regulator; it does not insert a positive activity-dependent mass
into \(|z+\lambda|^2\). The factor \(\eta/\epsilon\) cannot be omitted.

%%%%%%%%%%%%%%%%
\subsection{Zero-regulator limit and spectral stability}
\label{app:c-stability}
%%%%%%%%%%%%%%

For \(\alpha_0>0\) and \(\alpha_0^2K(0;z)<1\), the solution satisfies
\(\eta/\epsilon\to[1-\alpha_0^2K(0;z)]^{-1}\), giving
\begin{equation}
 J(z)=\lim_{\epsilon\downarrow0}J_\epsilon(z)
       =\frac{K(0;z)}{1-\alpha_0^2 K(0;z)}.
 \label{eq:c-new-offsupport}
\end{equation}
This is the squared resolvent outside the limiting eigenvalue support. For \(\alpha_0=0\) the unperturbed trace is simply
\(K(0;z)\) wherever that integral is finite.

If \(\alpha_0^2K(0;z)>1\), \(\eta\) instead approaches \(\eta_*(z)>0\)
specified by \(\alpha_0^2K(\eta_*;z)=1\), and
\begin{equation}
 J_\epsilon(z)\sim \frac{\eta_*(z)}{\alpha_0^2\epsilon}.
 \label{eq:c-new-interior}
\end{equation}
At a regular boundary point separated from the passive spectrum,
\(\alpha_0^2K(0;z)=1\) and
\(L(z)=\int\rho_0(\lambda)|z+\lambda|^{-4}d\lambda<\infty\).
Then \(\epsilon\sim\alpha_0^2L(z)\eta^3\), so that
\(J_\epsilon(z)\sim
 \alpha_0^{-2}[\alpha_0^2L(z)]^{-1/3}\epsilon^{-2/3}\).
This is a regulator asymptote at the boundary, not an activity power law.

The support criterion for a normal deformation by a Ginibre matrix agrees
with this branch change (Refs.~\cite{AFM2015,HoZhong2023}). For
\(\operatorname{Re}z\geq0\), the kernel at zero regulator is maximal at
\(z=0\), because each passive eigenvalue is nonnegative. Along the
positive real axis it decreases; at any fixed real part, a nonzero
imaginary part also decreases it. Thus, with a gapped passive spectrum,
the limiting bulk support first meets the imaginary axis at the origin.
For \(q<1\) the passive inverse-square moment is
\begin{equation}
 I_0=K(0;0)=(1-q)^{-3}.
 \label{eq:c-new-mp-m2}
\end{equation}
Indeed \(m(s)=\int\rho_0(\lambda)/(\lambda+s)\,d\lambda\) satisfies
\(qs m^2+(s+1-q)m-1=0\); evaluating at zero and differentiating yields
\(-m'(0)=(1-q)^{-3}\).
Consequently
\begin{equation}
 \alpha_c=(1-q)^{3/2},\qquad
 J(0)=\frac{(1-q)^{-3}}{1-(\alpha_0/\alpha_c)^2},
 \quad 0\leq\alpha_0<\alpha_c .
 \label{eq:c-new-static-threshold}
\end{equation}
All couplings are in units of the passive spectral scale. At \(q=1\),
\(K(0;0)\) diverges and the bulk support extends into the right half
plane for every \(\alpha_0>0\). Full Ginibre activity therefore does not
regularize a stable isostatic response in the unshifted model.

A finite sample can have an invertible \(\mathcal M\) even if it is
dynamically unstable. The formal inverse-trace is then not a
long-time response to constant forcing. Conversely the bulk criterion
does not guarantee stability for each finite sample; the actual
spectral abscissa must be checked. All zero-regulator limits here are
thermodynamic-first. In particular, Eq.~\eqref{eq:c-new-interior}
does not by itself determine a finite-size hard-edge distribution
or its moments. At strictly positive activity the real Gaussian
finite-\(N\) ensemble has support arbitrarily close to a regular
rank-deficient matrix; the distance to the singular set can have a smooth, nonzero density
down to zero. At \(z=0\) the singular set is a real hypersurface
near a rank-\(N-1\) matrix, and integrating the inverse square
of its transverse distance gives an infinite finite-\(N\) disorder
mean. Use a positive regulator or quantiles for finite-size
comparisons.

%%%%%%%%%%%%%%
\subsection{Independent weak-activity check}
\label{app:c-check}
%%%%%%%%%%%%%%%%%

For a fixed positive-definite \(\mathcal H_0\), put \(P=\mathcal H_0^{-1}\)
and \(m_k=\operatorname{tr}_N P^k\). Directly expanding
\((\mathcal H_0-\alpha_0\mathcal A)^{-1}\) before taking Gaussian
contractions gives the coefficient of \(\alpha_0^2\) in the
normalized squared inverse as
\begin{equation}
 \operatorname{tr}_N\mathbb E(P\mathcal A P^2\mathcal A^TP)
 +2\operatorname{tr}_N\mathbb E(P\mathcal A P\mathcal A P^2)
 =m_2^2+\frac{2m_4}{N}.
 \label{eq:c-new-weak-check}
\end{equation}
Here the equality concerns the averaged expansion coefficient at
\(\alpha_0=0\), not an expansion of an unregularized
finite-\(N\) disorder mean at nonzero activity.
The leading positive coefficient \(I_0^2\) agrees with
Eq.~\eqref{eq:c-new-offsupport}. This gives a check independently of the
Dyson closure and of how the regulator is removed.

%%%%%%%%%%%%%%%%%%%
\section{Linear susceptibility to an isotropic test force}
\label{app:compliance_derivation}
%%%%%%%%%%%%%%

Let \(\bm u=\delta\bm r\) be the displacement on the retained
subspace. For a fixed realization of the stability operator,
the deterministic response to a test force obeys
\begin{equation}
 \zeta\dot{\bm u}(t)=\mathcal M\bm u(t)+\bm f(t).
\end{equation}
This is also the equation for the probe-induced change in the mean
displacement in the presence of independent additive zero-mean noise.
When all eigenvalues of \(\mathcal M\) have negative real parts, the
homogeneous response decays. For a harmonic force with convention
\(\bm f(t)=\operatorname{Re}[\widehat{\bm f}\,e^{i\omega t}]\), the
steady complex displacement amplitude is
\begin{equation}
 \widehat{\bm u}=B(i\zeta\omega)^{-1}\widehat{\bm f},\qquad
 B(z)=z\mathbb I_{N_c}-\mathcal M.
 \label{eq:D-probe}
\end{equation}
Thus the physical frequency corresponds to \(z=i\zeta\omega\), not
to the Hermitization regulator \(\epsilon\). For a constant force the
long-time displacement is \(\bm u=-\mathcal M^{-1}\bm f\). If
\(\mathcal M\) is invertible but unstable, this expression specifies
a formal force-balanced displacement, not an attracting response.

For an ensemble of static test forces at fixed \(\mathcal M\), take
\begin{equation}
 \langle\bm f\rangle_f=0,\qquad
 \langle\bm f\bm f^\dagger\rangle_f
   =\sigma_f^2\mathbb I_{N_c},
 \label{FDT}
\end{equation}
where the subscript denotes the force average, not a matrix-disorder
average. This is a choice of forcing covariance, not a
fluctuation--dissipation relation. The displacement covariance is
\begin{equation}
 \langle\bm u\bm u^\dagger\rangle_f
   =\sigma_f^2\mathcal M^{-1}(\mathcal M^\dagger)^{-1}
   =\sigma_f^2(\mathcal M^\dagger\mathcal M)^{-1}.
 \label{eq:D-displacement-covariance}
\end{equation}
The order of the two factors is essential for the covariance matrix.
For the scalar trace, the two orders give the same result because they
have the same singular-value spectrum:
\begin{equation}
\begin{aligned}
 J
 &=\frac{1}{N_c}\operatorname{Tr}
             (\mathcal M^\dagger\mathcal M)^{-1}=\frac{1}{N_c}\operatorname{Tr}
             (\mathcal M\mathcal M^\dagger)^{-1}
  =\frac{1}{N_c}\sum_{k=1}^{N_c}\sigma_k^{-2}.
\end{aligned}
 \label{D11}
\end{equation}
Here \(\sigma_k\) are the singular values of \(\mathcal M\).
Therefore,
\begin{equation}
 \frac{\langle\|\bm u\|^2\rangle_f}{N_c}=\sigma_f^2 J .
 \label{D12}
\end{equation}
Regularizing the scalar trace gives the observable used in
Appendix~\ref{app:corrected-hermitization}.

The same computation for isotropic complex force amplitudes in
Eq.~\eqref{eq:D-probe} gives a squared response
\(J_{\rm ac}(\omega)=N_c^{-1}\operatorname{Tr}
 [B(i\zeta\omega)^\dagger B(i\zeta\omega)]^{-1}\).
A real harmonic displacement has an additional factor of \(1/2\) in
its time-averaged mean square at nonzero frequency. This response is
not the stationary variance under temporal noise: for example, a
white force with covariance \(2D_f\,\delta(t-t')\mathbb I_{N_c}\)
gives a variance per coordinate
\(D_f\pi^{-1}\int_{-\infty}^{\infty}J_{\rm ac}(\omega)\,d\omega\)
for a stable realization. Nor is \(J^{-1}\) a shear or bulk modulus:
such moduli require a specified strain or stress protocol and the
corresponding projection of the microscopic response.

%%%%%%%%%%%%%%%%%%%%%
\section{Contact counting and an optional passive shift}
\label{SubSec:Appl_RMT}

\subsection{Passive compatibility and the Wishart surrogate}
%%%%%%%%%%%%%%%%%%%%%%%%%

In a fixed frictionless contact network, each contact supplies one
scalar linearized normal-distance constraint. In the load-bearing
network, let \(N_{\rm b}\) be the number of particles after removing
rattlers and let \(Z\) be their mean contact coordination. Then
\(M_c=N_{\rm b}Z/2\) counts distinct contacts. With fixed cell shape,
\(N_c\) is the number of retained translational coordinates after
removing prescribed constraints and exact symmetry motions.
Thus \(N_c\simeq DN_{\rm b}\) in the large-system count, giving
Eq.~\eqref{eq:qZ}. The raw contact count is not a rank statement:
if \(\mathcal C\) maps the retained displacements to the changes of
contact lengths, the numbers of remaining zero modes and states of self-stress obey \(N_{\rm zero}-N_{\rm ss}=N_c-M_c\) by rank--nullity.

The longitudinal stiffness term is
\(\mathcal C^T\mathcal K\mathcal C\), where \(\mathcal K\) contains
the contact stiffnesses. It motivates the Wishart surrogate
\(\mathcal H_W=\mathcal R\mathcal R^T\) of
Eq.~\eqref{eq:wishart_def}
\cite{Franz2015,Ikeda2020,Ikeda22,Hayakawa2026}. The microscopic
compatibility matrix is sparse and structured; the surrogate replaces it with a dense random matrix to isolate the role of the aspect ratio.
Its Marchenko--Pastur law does not follow from contact counting alone.

%%%%%%%%%%%%%
\subsection{Model of an isotropic prestress shift}
%%%%%%%%%%%

The microscopic Hessian also contains geometric, force-dependent
terms. For instance, an interacting central-force pair contributes
a transverse curvature proportional to \(V_c'(r)/r\), which is
negative for a repulsive compressed contact. To illustrate such
softening in the surrogate, one may introduce a scalar shift in the
same passive spectral units as in the main text,
\begin{equation}
 \mathcal H_{\mathfrak e}=\mathcal H_W-\mathfrak e\mathbb I_{N_c},
 \qquad 0\leq\mathfrak e<\lambda_-,
 \qquad \lambda_-=(1-\sqrt q)^2 .
 \label{def:prestress}
\end{equation}
This is a phenomenological approximation, not a derivation of a
uniform microscopic prestress or of \(\mathfrak e(\phi)\). It is not
used in the figures or in the unshifted results of the main text.
For a stable passive bulk the lower edge must remain positive,
i.e., \(\lambda_- -\mathfrak e>0\). Setting
\(\lambda=x-\mathfrak e\) with \(x\) a bare Wishart eigenvalue gives
\begin{equation}
 \rho_{\mathfrak e}(\lambda)\,d\lambda
   =\rho_{\rm MP}(x)\,dx .
 \label{G7}
\end{equation}
The shifted inverse-square moment in
Appendix~\ref{app:c-stability} is then
\begin{equation}
 K_{\mathfrak e}=
 \int_{\lambda_-}^{\lambda_+}
  \frac{\rho_{\rm MP}(x)}{(x-\mathfrak e)^2}\,dx
  =-m_q'(-\mathfrak e),
 \label{eq:E-shift-kernel}
\end{equation}
where \(m_q(s)=\int \rho_{\rm MP}(x)/(x+s)\,dx\) over the bare
Wishart band is the transform used in Appendix~\ref{app:c-stability}.
Consequently, the limiting bulk threshold is \(\alpha_{c,\mathfrak e}=K_{\mathfrak e}^{-1/2}\), and the thermodynamic-first static response on its stable side is \(K_{\mathfrak e}/(1-\alpha_0^2K_{\mathfrak e})\).
At \(\mathfrak e=0\) these reduce to Eq.~\eqref{eq:c-new-static-threshold}. 
For \(q<1\), bringing the shifted lower edge to zero instead produces the shifted square-root onset of a regular Marchenko--Pastur edge, not the inverse-square-root density of the unshifted \(q=1\) spectrum.

The passive shift changes the stability threshold but does not change
the assumptions behind the active ensemble. In particular, neither
contact counting nor a change to the passive normal-mode basis
justifies independent Ginibre entries.

%%%%%%%%%%%%%%%%%%%%%%%%%%%%%%%%%%%

\end{document}